\documentstyle[aaspp4,flushrt]{article}

\lefthead{Knigge, C. et al.}
\righthead{A Self-Occulting Accretion Disk in the SW~Sex Star DW~UMa}

\begin{document}

\title{A Self-Occulting Accretion Disk in the SW~Sex Star DW~UMa
\footnote{Based on observations with the NASA/ESA Hubble Space
Telescope, obtained at the Space Telescope Science Institute, which is
operated by the Association of Universities for Research in Astronomy,
Inc. under NASA contract No. NAS5-26555.}}

\author{Christian Knigge
\footnote{Hubble Fellow}}
\affil{Department of Astronomy, 
Columbia University,
550 West 120th Street,
New York, NY 10027, USA} 
\authoremail{christian@astro.phys.columbia.edu} 

\author{Knox S. Long}
\affil{Space Telescope Science Institute, 3700 San Martin Drive,
Baltimore, MD 21218, USA}
\authoremail{long@stsci.edu} 

\author{D. W. Hoard}
\affil{Cerro Tololo Inter-American Observatory, Casilla 603, La
Serena, Chile}
\authoremail{dhoard@noao.edu}

\author{Paula Szkody}
\affil{Astronomy Department, University of Washington, Seattle, WA
98195, USA}
\authoremail{szkody@astro.washington.edu}

\and

\author{V. S. Dhillon}
\affil{Department of Physics and Astronomy, University of Sheffield,
Sheffield S3 7RH, UK}
\authoremail{vik.dhillon@sheffield.ac.uk}

\begin{abstract}

We present the ultraviolet spectrum of the SW~Sex star and
nova-like variable DW~UMa in an optical low state, as
observed with the {\em Space Telescope Imaging Spectrograph} on
board the {\em Hubble Space Telescope} (HST). The data are well
described by a synthetic white dwarf (WD) spectrum with $T_{eff} =
46,000 \pm 1000$~K, $\log{g} =  7.60 \pm 0.15$, $v\sin{i} = 370 \pm
100$~km~s$^{-1}$ and $Z/Z_{\odot} = 0.47 \pm 0.15$. For this 
combination of $T_{eff}$ and $\log{g}$, WD models predict $M_{WD} =
0.48 \pm 0.06$~M$_{\odot}$ and $R_{WD} = (1.27 \pm 0.18) \times
10^9$~cm. Combining the radius estimate with the normalization of the
spectral fit, we obtain a distance estimate of $d = 830 \pm 150$~pc.

During our observations, DW~UMa was approximately 3 magnitudes fainter
in V than in the high state. A comparison of our low-state HST 
spectrum to a high-state spectrum obtained with the {\em International
Ultraviolet Explorer} shows that the former is much bluer
and has a higher continuum level shortward of 1450~\AA. Since
DW~UMa is an eclipsing system, this suggests that an optically 
thick accretion disk rim blocks our view of the WD primary in the 
high state. If self-occulting accretion disks are common among the
SW~Sex stars, we can account for (i) the preference for
high-inclination systems within the class and (ii) their 
V-shaped continuum eclipses. Moreover, even though the emission 
lines produced by a self-obscured disk are generally still
double-peaked, they are weaker and narrower than those produced 
by an unobscured disk. This may allow a secondary line emission
mechanism to dominate and produce the single-peaked, optical lines 
that are a distinguishing characteristic of the SW~Sex~stars.

\end{abstract}


\section{Introduction}
\label{introduction}

The SW~Sex stars are a sub-class of nova-like cataclysmic variables
(NLCVs). Like all NLCVs, they are semi-detached, close binary  
systems, in which a late-type main sequence secondary star loses mass
to a WD primary via Roche-lobe overflow onto an 
optically thick accretion disk. However, they also exhibit a number of 
peculiar characteristics that are not shared by ``ordinary'' NLCVs 
(e.g. Thorstensen et al. 1991; Dhillon, Jones \& Marsh 1994; Rutten,
van Paradijs \& Tinbergen 1992): (i) 7 of the 10 current members of
the class are eclipsing systems (for an up-to-date list with
references, see Hoard, Thorstensen \& Szkody 2000); (ii) their
continuum eclipses appear to be more V-shaped (as opposed to
U-shaped) than those of other NLCVs; (iii) their optical emission
lines are single-peaked (instead of double-peaked, as expected for
lines formed in a rotating accretion disk); (iv) their low-excitation
optical emission lines are only weakly eclipsed at primary minimum,
but appear to undergo absorption events at the opposite orbital phase;
(v) the radial velocity curves derived from their optical emission
lines exhibit significant phase shifts relative to photometric
minimum. The V-shaped eclipses noted in (ii) are often interpreted as
evidence for flatter-than-expected radial temperature distributions
and non-stationary accretion in SW~Sex star disks (e.g. Rutten et
al. 1992; Horne 1999).

At least four different models have been proposed to explain the
SW~Sex stars' ``mysterious, but consistent behavior'' (Thorstensen et
al. 1991). The physical mechanisms invoked in these models are: (i) a
WD-anchored magnetic field that dominates the geometry of the
accretion flow (Williams 1989; Dhillon, Marsh \& Jones 1991); (ii) an
accretion 
disk wind that alters the radiative transfer of the optical emission
lines and the energy budget of the accretion disk (Honeycutt, Schlegel 
\& Kaitchuck 1986; Murray \& Chiang 1996; Dhillon \& Rutten 1995); (iii)
an accretion stream from the secondary star that overflows the
outer disk edge (Hellier \& Robinson 1994), possibly in combination
with either non-axisymmetric, vertical disk structure (Hoard 1998;
Hoard et al. 2000) or a flared, but not self-occulting disk (Hellier
1998); (iv) a disk-anchored magnetic field that acts as a
propeller and ejects a significant amount of material out of the
system (Horne 1999).

No real consensus has emerged to date regarding which, if any, of
these models is correct. In our view, this is partly because the
models have been tested almost exclusively against the 
known, optical characteristics of the SW~Sex stars -- which they 
were designed to reproduce in the first place. A promising, 
complementary approach is to compare the models to observation
in a qualitatively different observational regime. With this idea 
in mind, we have obtained high time- and medium 
spectral resolution ultraviolet (UV) spectroscopy of the SW~Sex star
DW~UMa with the {\em Space Telescope Imaging Spectrograph} (STIS) on
board the HST. Much to our initial 
disappointment, our observations found DW~UMa in a deep (V $\simeq$
17.6) low state (in the high state, V $\simeq$ 14.5). However, 
a comparison of our HST data to {\em International Ultraviolet
Explorer} (IUE) observations of DW~UMa in its ``normal'', high 
state reveals a surprising result: DW~UMa's far-UV continuum
(shortward of 1450~\AA) is {\em higher} in the low state than in 
the high state. As we will discuss, the simplest interpretation of 
this result is that the WD in DW~UMa is geometrically occulted by 
the rim of the accretion disk in the high state.

\section{Observations and Analysis}

The HST/STIS observations of DW~UMa took place on January 25 of 1999 (UT) 
and covered just over two complete cycles of DW~UMa's 3.28~hr
orbital period. Near-simultaneous optical photometry obtained from the 
MDM observatory puts DW~UMa at about $V\simeq 17.6$ around the time of 
the HST observations. The instrumental set-up consisted of the
52\arcsec~$\times$~0.2\arcsec~slit, the FUV-MAMA detectors and the
G140L grating, yielding a wavelength coverage of 1150~\AA~-~1720~\AA~
at a resolution of $\simeq$1~\AA~(FWHM). TIME-TAG mode was used
throughout, i.e. individual photon arrival times were recorded at a
sampling rate of 125$\mu$s. The data were split into suitable
``sub-exposures'' with the IRAF task INTTAG within
STSDAS/CALSTIS. These sub-exposures were then calibrated via the
standard CALSTIS pipeline, using the reference files available in
December 1999. The absolute flux scale of the flux-calibrated spectra 
is accurate to about 4\%.

In Figure~1, we show the out-of-eclipse UV spectrum of DW~UMa between 
HJD 2451204.122 -- 2451204.168, a time interval during which the flux
from the target was particularly stable. The corresponding orbital
phase range is 0.32 -- 0.65. The spectrum is quite blue and  dominated
by absorption lines, as expected for a hot WD. The thick line in
Figure~1 is the best-fitting synthetic WD spectrum, which was obtained
by interpolation from a grid of LTE model 
spectra calculated with TLUSTY/SYNSPEC (Hubeny, Lanz \& Jeffery 1994
Hubeny 1988; Hubeny \& Lanz 1995). The model spectrum is characterized
by seven free parameters: (i) the effective temperature 
($T_{eff}$); (ii) the surface gravity ($\log{g}$); (iii) the projected
rotational velocity ($v \sin{i}$); (iv) the metal abundance
($Z$); (v) the interstellar reddening ($E_{B-V}$); (vi) the interstellar 
neutral hydrogen column ($\log{N_H}$); (vii) the normalization
($N$). The Helium abundance was kept fixed at its solar value to match
the depth of the He~{\sc ii}~1640~\AA~feature.

Obvious/suspected interstellar lines were masked out in the fit,
except for the interstellar contribution to Ly $\alpha$, which was
modeled as a fully damped absorption profile. We also 
masked out three additional features, which we tentatively identify as  
blue-shifted components of C~{\sc iii}~1175~\AA, Si~{\sc
iv}~1397~\AA~and C~{\sc iv}~1549~\AA~(see bottom panel of
Figure~1). All of these features lie at $-1700 \pm 150$~km~s$^{-1}$
from the rest wavelengths of their parent lines and are therefore
probably formed in an outflow from the system. This identification is
also supported by (i) the P-Cygni character of the fit residuals in the
vicinity of the two Carbon lines (especially C~{\sc iv}~1549~\AA), and
(ii) the presence of {\em two} features in the residuals shortward of
1400~\AA, whose spacing is consistent with the two components of the
Si~{\sc iv}~1397~\AA~doublet. 

The fit shown in Figure~1 gives a $\chi^2$ of 1079 for 712 degrees of
freedom ($\chi^2_v=1.5$). The corresponding model parameters are
listed in Table~1. The low reddening and small Hydrogen column we
derive are consistent with each other, with a previous reddening 
determination (Szkody 1987) and with the total galactic value of 
$\log{N_H} = 19.8$ in the direction of DW~UMa (Dickey \& Lockman
1990). Table~1 also contains estimates of the mass and radius of
DW~UMa's WD, which are based on the carbon-core WD models of Panei,
Althaus \& Benvenuto (2000) for our best-fit values for $T_{eff}$ and
$\log{g}$. The distance estimate in Table~1 is based on 
the normalization of the spectral fit and the WD radius estimate. The 
relatively large value of $d=830 \pm 150$~pc is consistent with
the lower limit of $450 - 850$~pc (depending on the spectral type of the
secondary) obtained by Marsh \& Dhillon (1997). Potential sources of
systematic error on our parameters include: (i) the possible
contribution of other components to the spectrum (e.g. the outflow or
a residual disk); (ii) the choice of masking windows in the spectral
fit; (iii) the application of a mass-radius relation for isolated WDs
to an accretion-heated WD in a NLCV; (iv) our assumption that the WD
is entirely unobscured.

\section{Discussion}

For comparison with our low state HST observations, we also show in
Figure~1 a high state, out-of-eclipse spectrum of DW~UMa obtained 
approximately 13 years earlier with IUE (SWP27097; Szkody 1987). The
IUE spectrum is much redder than the HST one, and, unlike the latter,
contains strong emission lines. Most importantly, however, the 
UV continuum shortward of 1450~\AA~is actually higher in the low state 
HST spectrum than in the high state IUE one. Thus if we are correct in
attributing the low-state UV light to the WD primary, our view of this 
system component must be blocked in the high state.
\footnote{The only alternative we can think of is that the WD is much
cooler in the high state than in the low state. We discard this as 
physically implausible, since the temperature of the WD is expected to
increase with increasing accretion rate (Sion 1999).} 
We have also inspected the five other (2 SWP and 3 LWP) UV 
spectra of DW UMa that were taken by IUE. All of them share the flat
UV continuum slope of the high-state IUE spectrum shown in Figure~1, and 
none exhibits a significantly higher UV flux level. Since the combined
IUE observations cover the entire orbital cycle (Szkody~1987), the 
WD in DW~UMa appears to be obscured at most or all binary phases
during the high state.

The simplest interpretation of these findings is that, in the 
high state, our view of the WD is blocked by the rim of a flaring,
optically thick accretion disk.
\footnote{A flat, geometrically thin, optically thick disk can occult
no more than half of the projected area of the WD. The presence of
such a disk can therefore reduce the high state WD flux by at most a
factor of two. The maximum discrepancy between low-state and high-state
continua is larger than this. Moreover, the hot, inner regions of a
geometrically thin disk would themselves contribute substantially to
the far-UV flux, offsetting the effect of the WD occultation.}
For a disk with semi-opening angle $\alpha$
(measured at its outer edge), this requires that $i+\alpha \geq
90^o$. In a deeply eclipsing system like DW~UMa, this condition may
well be met (see below). Our results can then be explained as
follows. In the high state, the inner disk, the WD and the front of
the disk are all hidden from view by the disk's front rim. The only
elements that remain visible are the cool, outer regions on the back
of the disk, as well as the occulting disk rim itself. 
The observed UV flux in the high state is therefore much lower (and
the UV spectrum much redder) than if the hot, inner disk, and the WD
at its center, were visible. In the low state, the mass transfer rate
from the secondary star is severely reduced or completely shut
off. The accretion disk then becomes optically thin (or disappears
completely), and the WD emerges. In order to be consistent
with the observations, this picture requires that the hot, compact WD
should match or exceed the far-UV -- but not the optical -- output of
the much cooler, visible regions of the high-state accretion
disk. This is plausible, given the differences in effective
temperature and projected area associated with these system
components. For example, for two blackbodies at temperatures
$T_1=45,000$~K and $T_2=15,000$~K, the ratio
$B_\lambda(T_1)/B_\lambda(T_2)$ is approximately 92 at
$\lambda=1450$~\AA~but only about 6 at $\lambda=5550$~\AA. So if the
visible portions of the high state disk have a characteristic
temperature near $T_2$ and present a projected area around 100 times
that of the WD, the UV flux in the low state will be similar to that
in the high state, whereas the optical flux will be 3 magnitudes
(15$\times$) lower, as is observed. 

Thus the simple hypothesis that DW~UMa in the high state harbors a
self-occulting accretion disk can account for: (i) the red UV colors 
in the high state; (ii) the blue UV colors in the low state; (iii) the
approximate equality of the high- and low-state UV flux levels; (iv)
the lack of such an equality in the optical waveband. Given this
success, it is interesting to explore the impact of  the
self-occultation scenario more generally, i.e. beyond its specific
application to our HST observations of DW~UMa.
First, it is obvious that the preference for eclipsing systems in the
SW~Sex class is explained quite naturally if at least some aspects of
the SW~Sex syndrome are related to the presence of a self-occulting
accretion disk. Second, Smak (1994) has pointed out that a flared
accretion disk will produce increasingly V-shaped 
continuum eclipse light curves as it is viewed increasingly
edge-on. This effect is due to the increasing relative contribution of
the facing disk rim to the total light. But an additional effect is at
work in a {\em self-occulting} disk: since the hot disk center is
permanently hidden from view, even the top of the disk produces a more
V-shaped eclipse in this case (Rutten 1998). Third, the line profiles
produced by a self-obscured, flared disk will be significantly
different from those produced by a fully visible disk. This may have a
bearing on the peculiar optical emission line behaviour
characteristics of the SW~Sex stars.

To illustrate and explore the last two points, we show in Figure~2
some synthetic eclipse light curves and line profiles. In calculating
these models, the following parameters were adopted: 
(i) primary mass: $M_{WD} = 0.5M_{\odot}$; 
(ii) mass ratio and inclination: $q=M_{2}/M_{WD} = 0.4$ and $i=82^o$;
(iii) secondary radius: $R_2=0.27 R_{\odot}$;
(iv) accretion disk radius: $R_{disk} = 0.8 L_1$;
(v) radial surface brightness distribution: $S \propto R^{-3/2}$.
The combination of $q$ and $i$ in (ii) gives an eclipse width 
roughly consistent with that of DW~UMa (Dhillon et al. 1994). $R_2$ is 
based on Smith \& Dhillon's (1998) mass-radius relation for CV
secondaries, taking $M_2 = q M_{WD}$. $L_1$ is the distance
from the WD to the inner Lagrangian point, and $R_{disk}$ is based on
DW~UMa's total eclipse width in the high state (Dhillon et al. 1994). The
radial surface brightness distribution in (v) is applied to both lines
and continuum. To explore the effect of the shape of the disk on line
profiles and continuum light curves, we have calculated two types of 
models. In one, the disk is described as flat, but with
a raised rim at its outer edge; in the other, the disk is taken to
have a constant aspect ratio H/R. These two shapes are the limiting
cases of concave disks. In both cases, we have set 
$\alpha=\arctan{[H(R_{disk})/R_{disk}]} = 10^o$. We have also considered two
limiting cases for emission from the facing rim: for each disk shape, 
we present models with and without rim emission. In the former, the
surface brightness of the rim is the same as at the top edge of the
disk. All model line profiles were calculated in both optically thin
and thick limits. For the latter, we generalized slightly the 
treatment given in Horne \& Marsh 1986 to account for the flare of the
disk.

The third row of panels in Figure~2 shows that self-obscured
disks do indeed produce much more V-shaped continuum eclipses than
a flat disk, regardless of the precise disk shape. As expected, the 
effect is particularly pronounced when the rim contributes
significantly to the 
emission. Thus self-occultation can easily account for the V-shaped
continuum eclipses of the SW~Sex stars and for the correspondingly 
flat T(R) distributions obtained in eclipse mapping studies of 
SW~Sex stars (Rutten et al. 1992). The bottom row of
panels in Figure~2 shows that the line profiles produced by
self-obscured disks are weaker and narrower than those 
produced by a flat disk. This is because the bright, quickly rotating 
inner disk regions are no longer visible in a self-occulted
disk. However, the line profiles are generally still double
peaked. Thus self-occultation alone can probably not account for 
the single-peaked profiles of the optical lines in the SW~Sex stars. 
Self-occultation may nevertheless be key to the production of
single-peaked line profiles in some of these systems. In particular,
the weakness of lines produced by a self-obscured disk may allow 
a secondary, normally negligible, line emission mechanism to dominate. 
This mechanism could actually be at work in all NLCVs, but may only be
observable in the SW~Sex stars.

\vspace*{0.1cm}

We are grateful to Joe Patterson, Jonathan Kemp, John Thorstensen and
Cindy Taylor for their contributions to this project and to Leandro
Althaus for providing his WD mass-radius relations in electronic form.
Support for this work was provided by NASA through grant  
number GO-7362  from the Space Telescope Science Institute (STScI),
which is operated by AURA, Inc.,under NASA contract
NAS5-26555. Additional support for CK was provided by NASA through
Hubble Fellowship grant HF01109.0198A, also awarded by STScI.

\bibliographystyle{apj}
\bibliography{bibliography}

\clearpage

\protect\begin{deluxetable}{lc}
\tablewidth{230pt}
\footnotesize
\tablecaption{Parameters inferred from the spectral fit.\label{tbl-wd}}
\tablehead{
\colhead{Parameter}& 
\colhead{Value\tablenotemark{a}}}
\startdata
$T_{eff}$~[K]& 46,000 $\pm$ 1,000~ \nl
$\log{g}$~[$\log{}$cm~s$^{-2}$] & 7.60 $\pm$ 0.15 \nl
$v\sin{i}$~[km~s$^{-1}$] & 370 $\pm$ 100 \nl
$Z$~[$Z_{\odot}$] & 0.47 $\pm$ 0.15 \nl
$N=4\pi (R_{WD}/d)^2$~[10$^{-24}$] & 3.08 $\pm$ 0.15\tablenotemark{b} \nl
$E_{B-V}$~[mag] & 0.005 $\pm$ 0.005 \nl
$\log{N_H}$~[$\log{}$cm$^{-2}$] & 19.4 $\pm$ 0.1~ \nl \nl
$M_{WD}$~[$M_{\odot}$] & 0.48 $\pm$ 0.06 \nl
$R_{WD}$~[$10^9$~cm] & 1.27 $\pm$ 0.18 \nl
d~[pc] & 830 $\pm$ 150\tablenotemark{c} \nl
\enddata
\tablenotetext{a}{Uncertainties on the fit parameters correspond to 
1-$\sigma$ confidence intervals for all parameters considered
jointly after scaling the errors on the data by
$\sqrt{\chi^2_v}=\sqrt{1.5}$.}
\tablenotetext{b}{This is the formal error of the fit. There is also a
4\% uncertainty in the absolute flux calibration 
of the spectrum and an uncertainty of up to 20\% in the normalization
of the models (c.f. Brown, Ferguson \& Davidsen 1996; Knigge et al. 1997)}
\tablenotetext{c}{The quoted uncertainty accounts for the formal
errors on $R_{WD}$ and $N$, as well as for the uncertainties in flux
calibration and model normalization.} 
\protect\end{deluxetable}

\clearpage

\begin{figure}
\figurenum{1}
\plotone{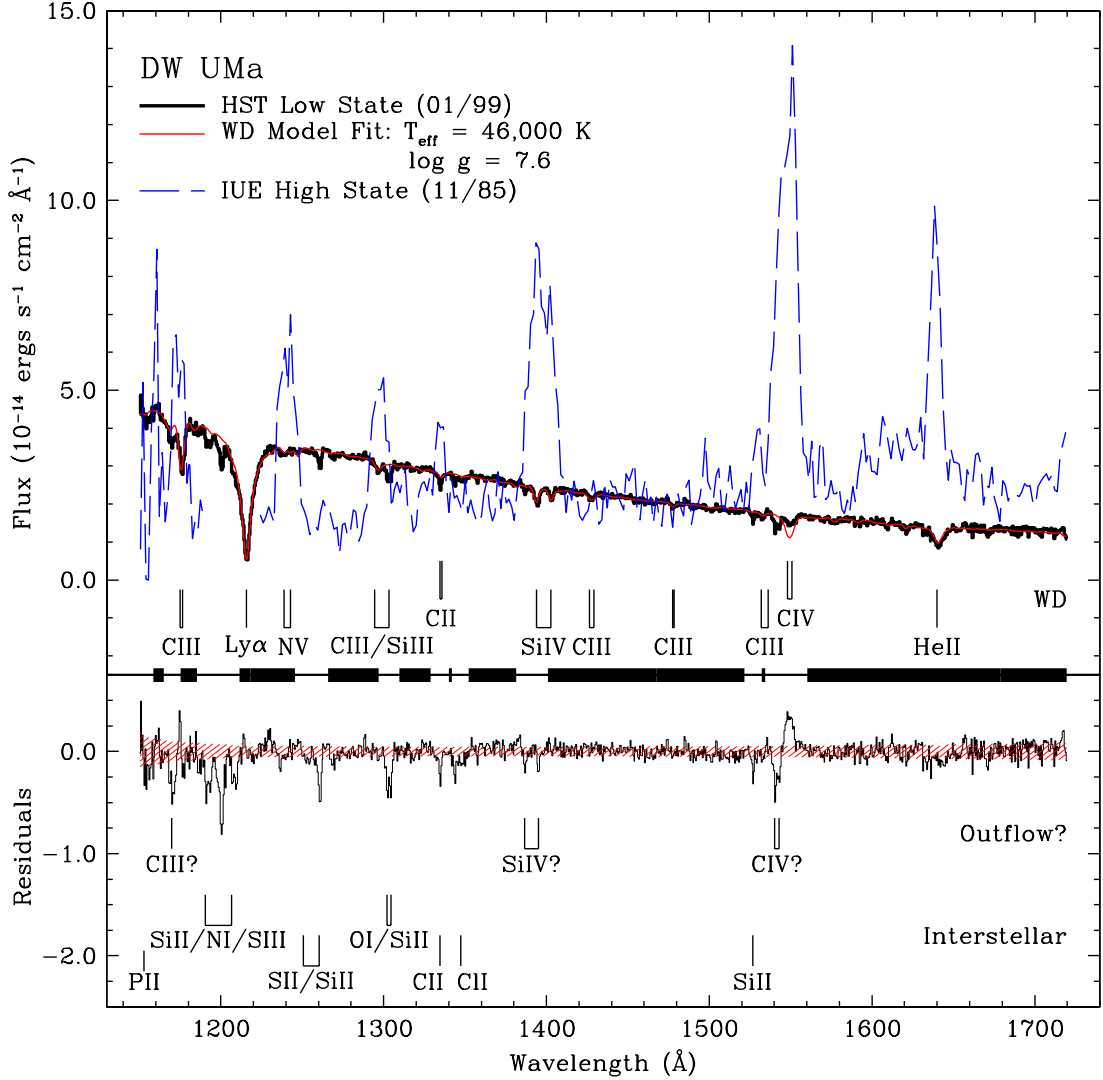}
\protect\vspace*{-4cm}
\figcaption{{\em Top Panel:} DW~UMa's out-of-eclipse
UV spectrum in the low state (thin line). Also shown are the best-fit
synthetic WD spectrum (thick line; see Table~1 for parameters) 
and the out-of-eclipse UV spectrum in the high state, as
observed with IUE (dotted line). The strongest transitions in the
WD-dominated low-state spectrum are indicated. {\em Bottom Panel:}
The residuals of the WD model fit to the low state spectrum 
(thin line); the shaded region encloses $\pm 1\sigma$. Suspected
interstellar lines are marked, as are three features that are probably 
formed in an outflow from the system (see text). The thick lines on
the axis separating the panels mark the fit windows.}
\end{figure}

\clearpage
\begin{figure}
\figurenum{2}
\plotone{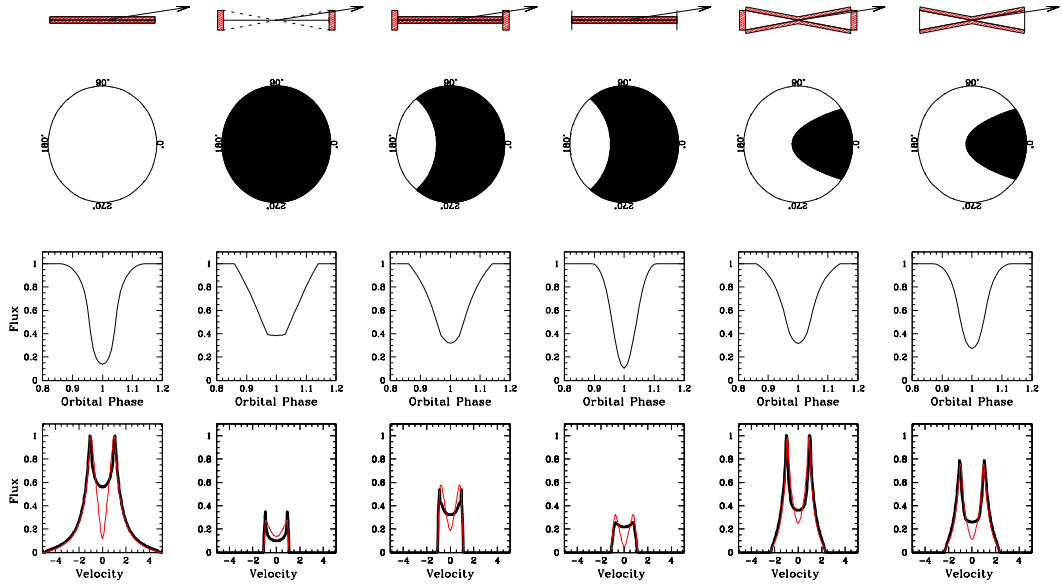}
\figcaption{{\em Top Row:} The disk/rim
configurations (roughly to scale) for which light curves and line
profiles have been calculated. From left to right, we consider (i) a
flat disk, (ii) an 
isolated rim, (iii) a self-obscured, flat disk with an emitting rim,
(iv) a self-obscured, flat disk with a dark rim, (v) a self-obscured,
flared disk with an emitting rim, (vi) a self-obscured, flared disk
with a dark rim. Shaded areas mark emitting regions 
(the thickness of emitting layers has been greatly exaggerated).
Arrows indicate the direction towards the observer. {\em Second
Row:} Top view of the 
model disks. Dark areas indicate self-occulted regions. {\em
Third Row:} Normalized continuum light curves produced by the 
disk models. {\em
Bottom Row:} Optically thin (solid lines) and optically thick (dotted
lines) out-of-eclipse line profiles produced by the disk models. The
unit of the ordinate axis is the projected velocity at the outer disk
edge. Line profiles in the left panel are shown scaled to a peak flux
of unity; fluxes in all other panels are relative to those
in the left panel. A 5-point boxcar smoothing has been applied to all
profiles (the dispersion is 0.034 velocity units).}
\end{figure}

\end{document}